\documentclass{article}
\usepackage{spconf,graphicx}
\usepackage{amsmath,amsfonts,amssymb}
\usepackage{textcomp}
\usepackage{color}
\usepackage{xcolor}
\usepackage{mathrsfs}
\usepackage{amsthm,algorithmicx}
\usepackage{algorithm,algpseudocode}
\usepackage{algpseudocode}
\usepackage{cite}
\usepackage{soul}
\usepackage{graphicx}
\usepackage{epsf}
\usepackage{bm}
\usepackage{multirow,booktabs}
\usepackage{latexsym}

\usepackage{hyperref}
\usepackage[export]{adjustbox}
\usepackage{tikz}
\usepackage{dsfont}
\usepackage[RPvoltages]{circuitikz}
\usetikzlibrary{arrows,shapes}
\usepackage{tkz-euclide}
\usetikzlibrary{backgrounds}
\usepackage{epstopdf} 
\DeclareGraphicsExtensions{.jpg,pdf,.png}
\RequirePackage{amsopn,mathcomSTEv4}

\theoremstyle{plain}
\newtheorem{theorem}{\textbf{Theorem}}

\theoremstyle{definition}
\newtheorem{definition}{\textbf{Definition}}
\theoremstyle{remark}

\DeclareGraphicsExtensions{.jpg,pdf,.png,.bmp}

\usetikzlibrary{positioning}
\usetikzlibrary{shapes,arrows}
\tikzstyle{block} = [draw, fill=white, rectangle, 
minimum height=2em, minimum width=3em]
\tikzstyle{input} = [coordinate]
\tikzstyle{output} = [coordinate]
\tikzstyle{pinstyle} = [pin edge={to-,t,black}]

\newcommand{\mlambda}{\boldsymbol{\Lambda}}

\definecolor{edit}{HTML}{1212EE}
\definecolor{unclear}{HTML}{AA6622}
\definecolor{suggest}{HTML}{55BB44}

\newcommand{\Hscr}{\mathscr{H}}
\newcommand{\rank}{{\rm rank}}

\title{Two-snapshot DOA  Estimation via Hankel-structured  Matrix Completion}
\name{Mohammad~Bokaei$^\dagger$, Saeed~Razavikia$^\dagger$, Arash~Amini$^\dagger$, and Stefano Rini$^\star$}
\address{ $^\dagger$Advanced Communication Research Institute (ACRI), Sharif University of Technology, Tehran, Iran\\
    $^\star$ Electrical and Computer Engineering department, National Chao Tung University, Hsinchu, Taiwan
}

\begin{document}
%\ninept
%
\maketitle
\begin{abstract}

In this paper, we study the problem of estimating the direction of arrival (DOA) using a sparsely sampled uniform linear array (ULA). Based on an initial incomplete ULA measurements, our strategy is to choose a sparse subset of array elements for measuring the next snapshot. Then, we use a Hankel-structured matrix completion to interpolate for the missing ULA measurements.
Finally, the source DOAs are estimated using a subspace method such as Prony on the fully recovered ULA. We theoretically provide a sufficient bound for the number of required samples (array elements) for perfect recovery. The numerical comparisons of the proposed method with existing techniques such as atomic-norm minimization and off-the-grid approaches confirm the superiority of the proposed method.

\end{abstract}
\begin{keywords}
%
%{\color{blue} SR2AA: needs fixing}{\color{green}  Is it Ok, now?}
%
Direction of arrival, matrix completion, non-unifrom sampling,  off-the-grid compressed sensing, super-resolution.
\end{keywords}
\section{Introduction}
\label{sec:intro}

Conventional methods for estimating the direction of arrival (DOA) of multiple sources, mainly rely on the auto-correlation function of the received data, which requires the availability of samples of multiple snapshots \cite{schmidt1982signal}.
Recent advances, however,  have shown that the samples of even a single snapshot could be sufficient for estimating DOAs if the sources are sparsely present in the field of view
\cite{jin2009joint}. 
To take advantage of the latter sparsity, it is common to use 
%The latter approach employs 
sparse recovery techniques in compressed sensing that require discretizing the range of angles (an angular grid). This comes with the drawback of grid mismatch error \cite{Chi2011mis} which cannot be completely eliminated \cite{yang2014discretization}.

%%%%%%%%%%%%%%%%%%%%%%%%%---Amini---%%%%%%%%%%%%%%%%%%%%%%%%%%%%%%%%
%The latter approach employs sparse recovery techniques in compressed sensing that require discretizing the range of angles (an angular grid). This comes with the drawback of grid mismatch; i.e., an error is introduced if the sources are not exactly placed on grid angles. 
%Although by increasing the resolution of the grid one might get more accurate DOA estimates with higher computational cost \cite{duarte2013spectral}, the mismatch error cannot be completely eliminated \cite{yang2014discretization}.
%%%%%%%%%%%%%%%%%%%%%%%%%%%%%%%%%%%%%%%%%%%%%%%%%%%%%%%%%%%%%%%%

A super-resolution method is introduced \cite{candes2014towards} for estimating a sparse mixture of single-frequency signals based on uniform samples but without assuming any grid; the only constraint is a minimum separation between the frequencies. The method can be translated into a grid-less sparse DOA estimation technique based on a uniform array which is not an efficient array for source localization  \cite{xu2014compressed}. 
For sparse linear arrays (SLA) \cite{tang2013compressed} proposed a grid-less sparse recovery method based on atomic-norm minimization (ANM). For perfect estimation, the method requires a minimum angular separation between the sources.

%A grid-less frequency estimation problem named the super-resolution technique is provided in \cite{candes2013super, candes2014towards}. While this method does not retrieve exact DOAs, a minimum angular separation between sources should be satisfied.
%More importantly, this method relies on uniform arrays which are not efficient for DOA estimation problem \cite{xu2014compressed}. 
%A grid-less sparse recovery method based on atomic norm minimization (ANM) was proposed in \cite{tang2013compressed} to determine DOAs using SLA. These approaches are accurate under a DOA separation constraint.

It is known that the Hankel matrix form of the samples of a ULA represents a low-rank matrix when the number of sources is small \cite{hua1992estimating}, which assists in estimating the DOAs. 
EMaC is a grid-less approach that extends this idea to a SLA \cite{chen2014robust,razavikia2019reconstruction,razavikia2019sampling}. The method essentially interpolates the sparse samples using matrix completion in order to achieve uniformly-spaced samples (SLA to ULA), and applies a standard super-resolution technique tailored for ULAs.  Compared to ANM, EMaC is shown to retrieve more sources with the same number of samples.

\begin{figure*}[t]
	\centering
	\scalebox{0.7}{
		\begin{tikzpicture}[scale=1, transform shape]
		\tikzstyle{block} = [draw, fill=blue, rectangle]
		\tikzstyle{input} = [coordinate]
		\tikzstyle{output} = [coordinate]
		\tikzstyle{multiply} = [draw, circle ]
		%%%%%%%%%%%%%%%%% Define Nodes %%%%%%%%%%%%%%%%%%%%%%%%%%%%%%%%%%%%%%%%%%%
		\node [name=Pattern,node distance=0] {\includegraphics[width = 1.4in]{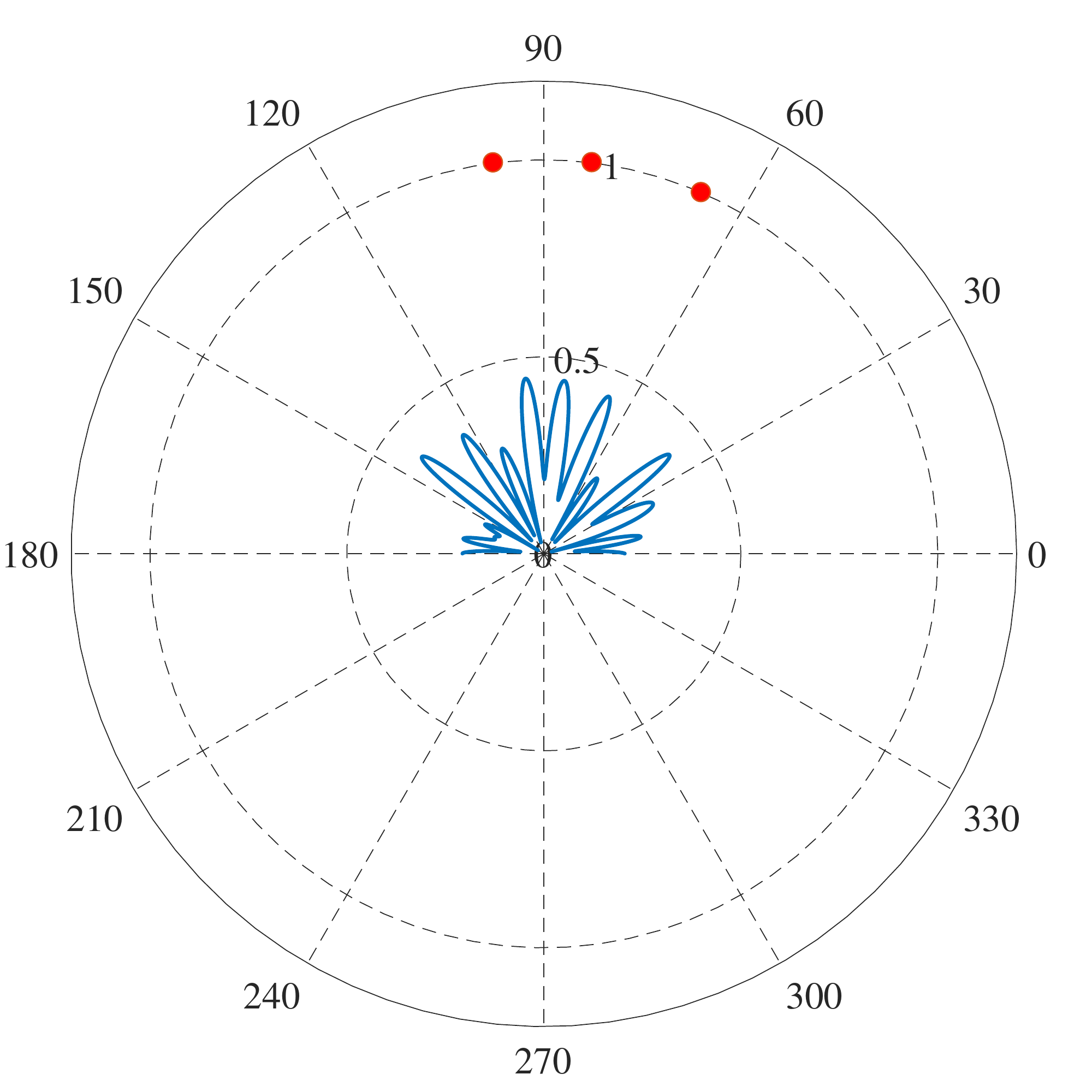}};
		\node [right of=Pattern, name=InputArray,node distance=3.5cm] {\includegraphics[width = 0.1in, height =1.2in]{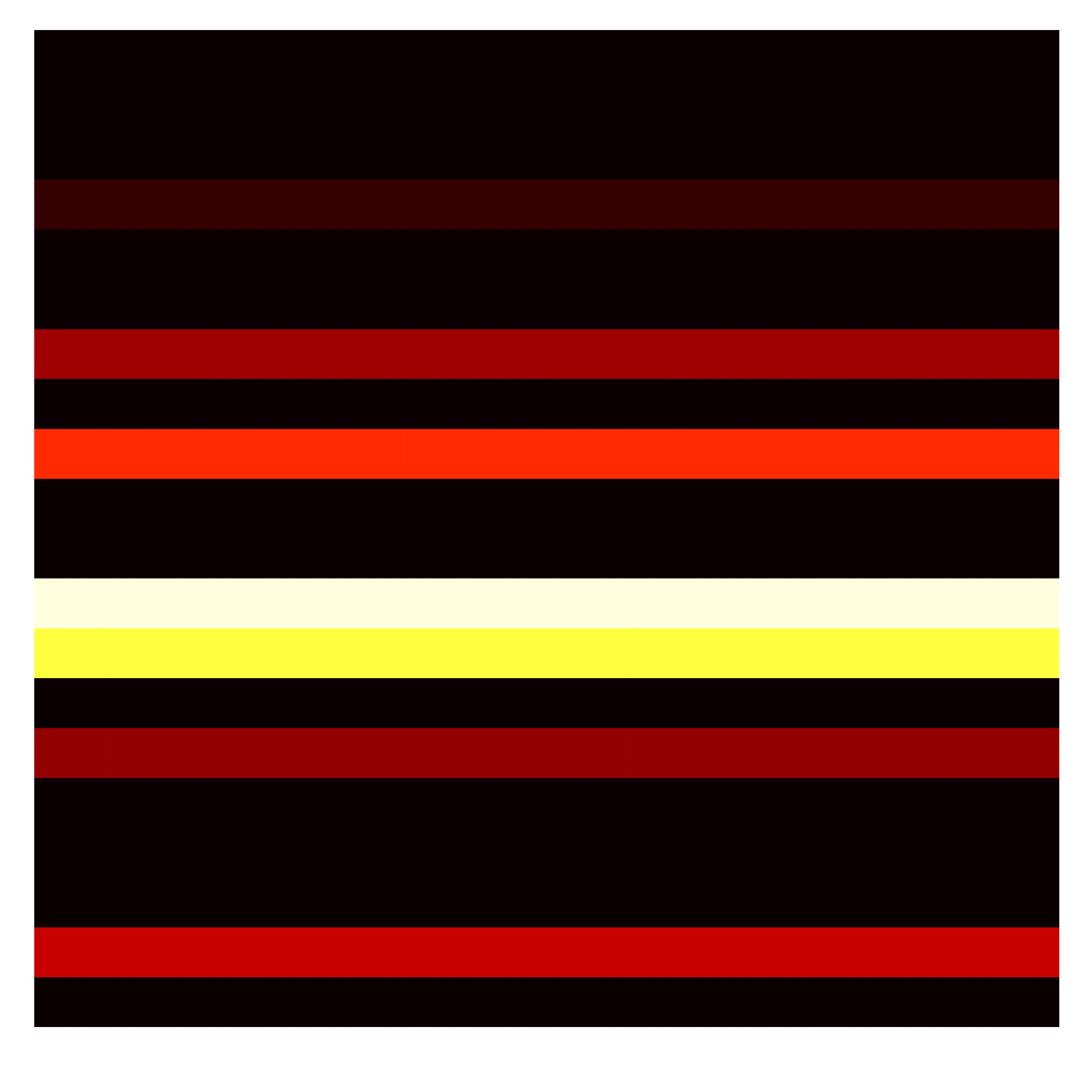}};
		\node [right of=InputArray, name=Hankel ,node distance=2.5cm] {\includegraphics[width = 1in, height=1in]{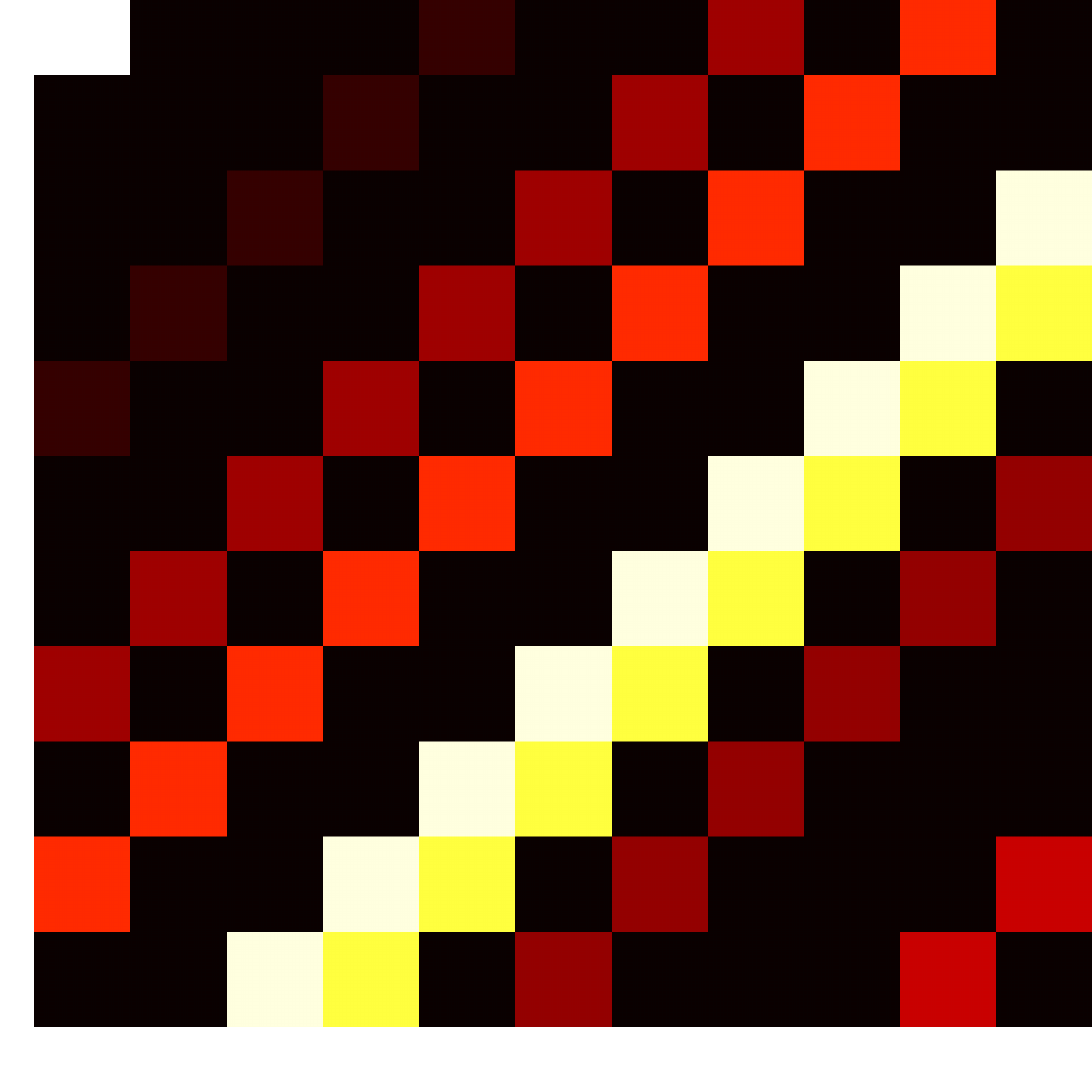}};
		%		\node [circle, draw=black,fill=blue!10, right of=Hankel, name=fisrstmultiply, node distance=3cm]{\large $\times$};
		%		\node [below of=fisrstmultiply,name = WeightedRow, node distance=2.5cm]  {\includegraphics[width = 1in, height= 0.1in]{figs/WeightRow.eps}};
		%		\node [right of=WeightedRow ,name = WeightedColumn, node distance=1.5cm]  {\includegraphics[width = 0.1in, height= 1in]{figs/WeightColumn.eps}};
		%		\node [right of=Hankel,name = WeightedInput, node distance=3.5cm]  {\includegraphics[width = 1in, height= 1in]{figs/weightedInputHankel.eps}};
		\node [rectangle,draw=black,rounded rectangle, fill=blue!20, right of=Hankel,name = Recovery, node distance=4cm]  {$\substack{\mathbf{Low~ Rank} \\\mathbf{Matrix ~ Completion}}$};
		%		\node [right of=Recovery, name=WeightedOutput, node distance=2.5cm] {\includegraphics[width = 1in, height= 1in ]{figs/recoverdWeightdHankel.eps}};
		%		\node [circle, draw=black,fill=blue!10, left of=WeightedOutput, name=secrstmultiply, node distance=3.5cm]{\large $\times$};
		\node [ right of=Recovery, name=HankelOutput, node distance=4cm] {\includegraphics[width = 1in, height= 1in ]{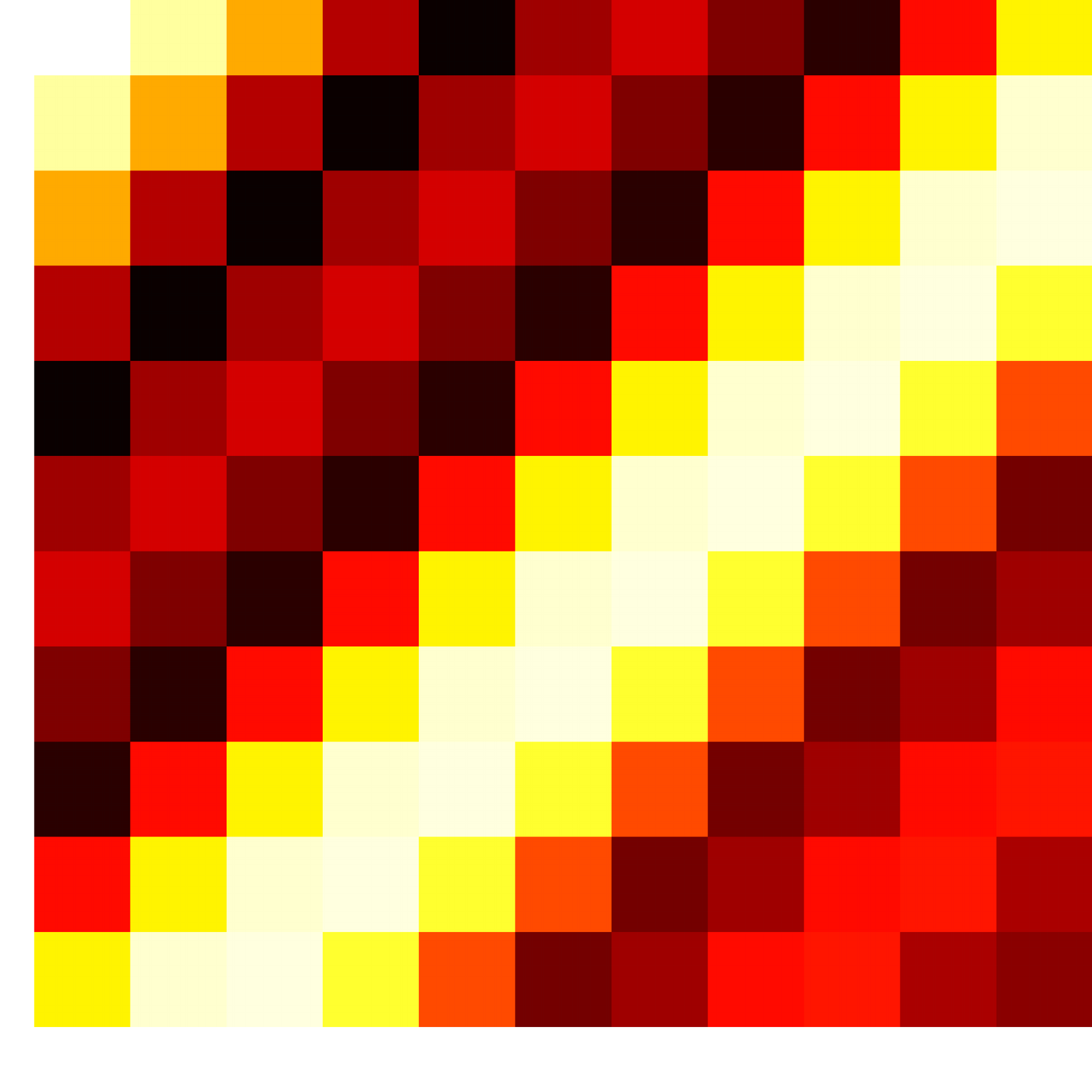}};
		\node [ right of=HankelOutput, name=ArrayOutput, node distance=2.5cm] {\includegraphics[width = 0.1in, height= 1.2in ]{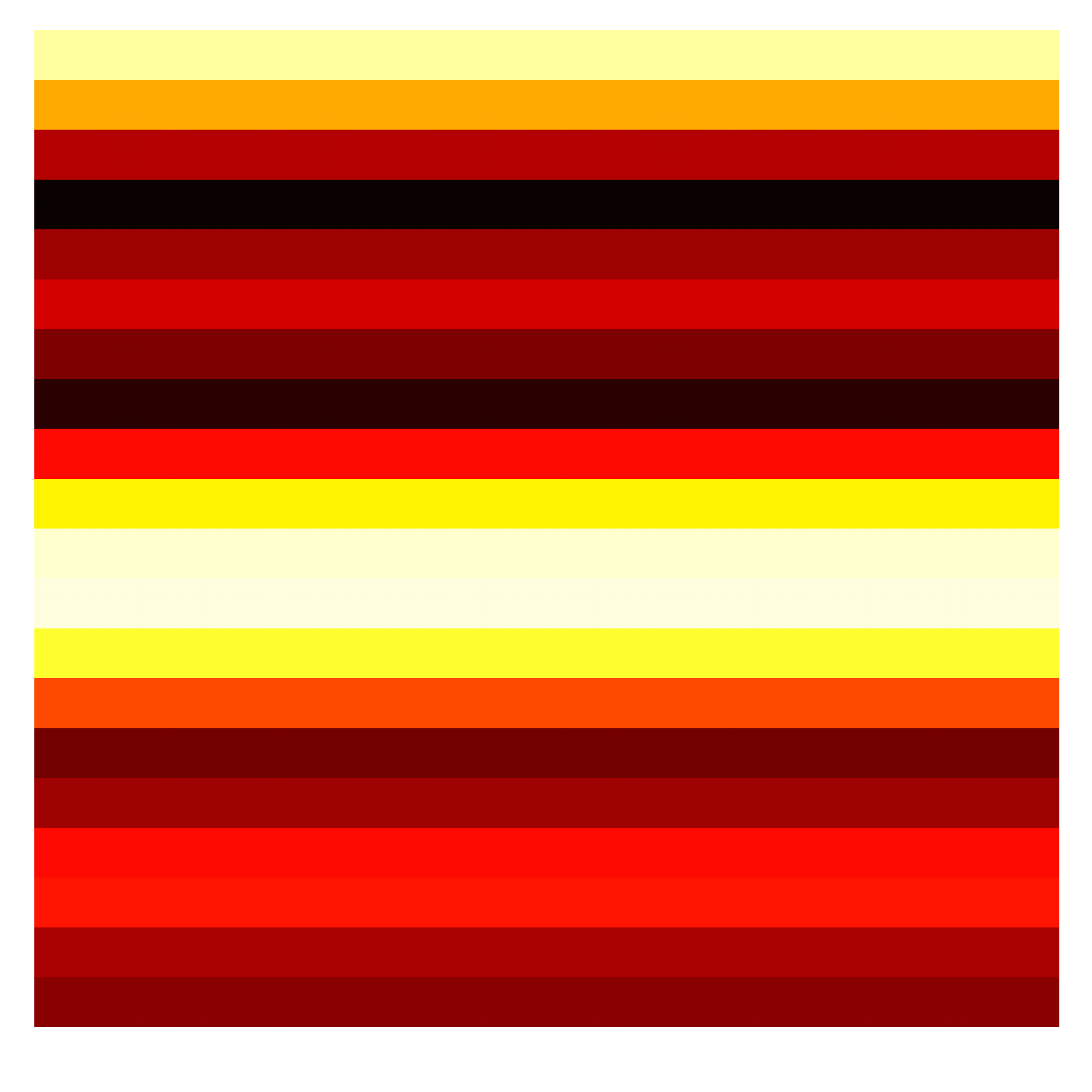}};
		\node [ right of=ArrayOutput, name=PatternOut, node distance=3.5cm] {\includegraphics[width = 1.4in ]{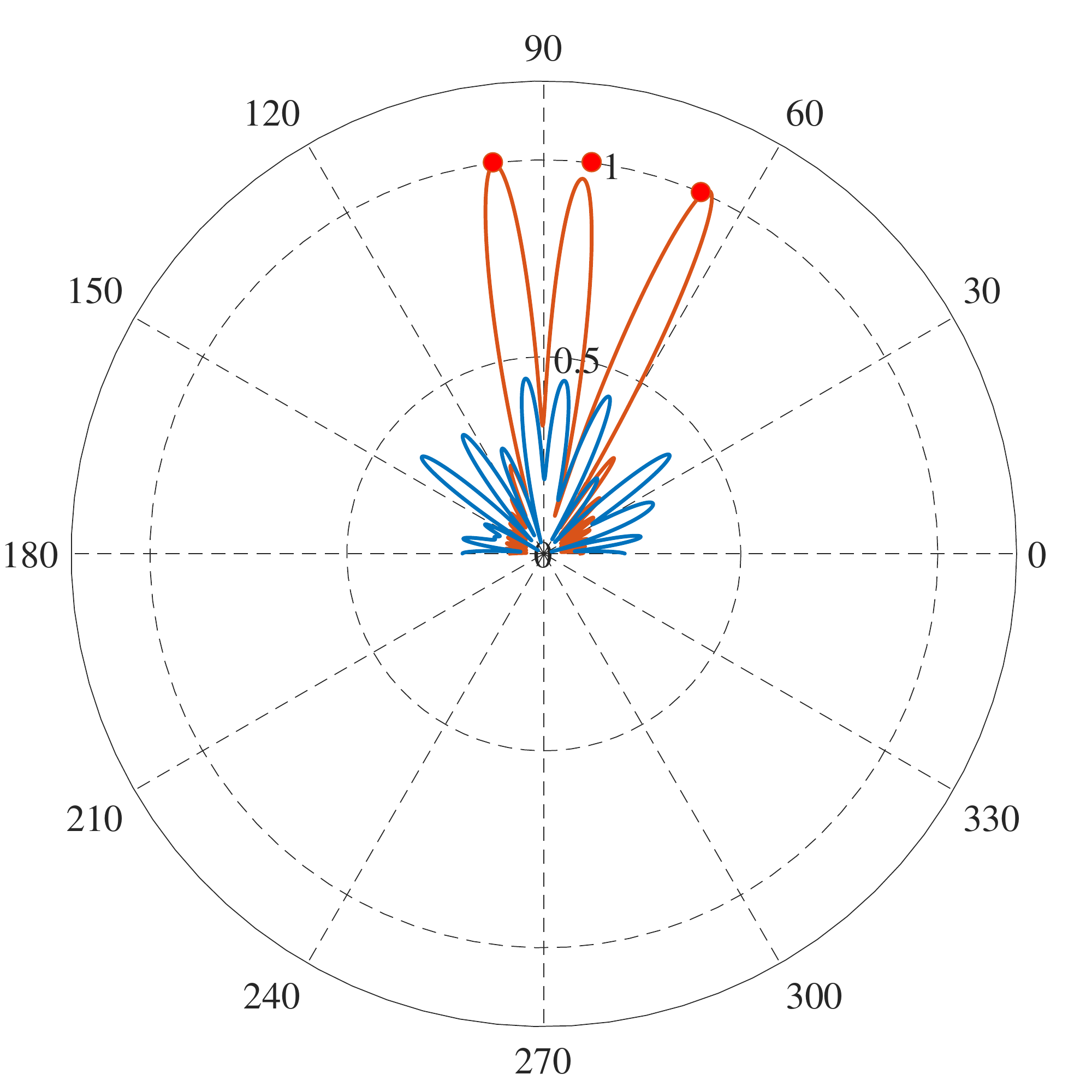}};
		%%%%%%%%%%%%%%%%% Connect the Nodes %%%%%%%%%%%%%%%%%%%%%%%%%%%%%%%%%%%%%%
		\draw [thick,->] (Pattern)  -- (InputArray);
		\draw [thick,->] (InputArray)  -- node[above,midway] {$\mathscr{H}$} (Hankel);
		\draw [thick,->] (Hankel)  -- (Recovery);
		%		\draw [thick,->] (9,-1.5)  -- (fisrstmultiply);
		%		\draw [thick,->] (9,-3.5)  -- node[right,midway] {$-1$} (secrstmultiply);
		%		\draw [thick,->] (fisrstmultiply)  -- (WeightedInput);
		%		\draw [thick,->] (WeightedInput)  --  (Recovery);
		\draw [thick,->] (Recovery)  --  (HankelOutput);
		%		\draw [thick,->] (WeightedOutput) -- (secrstmultiply);
		%		\draw [thick,->] (secrstmultiply) -- (HankelOutput);
		\draw [thick,->] (HankelOutput)  -- node[above,midway] {$\mathscr{H}^{\dagger}$} (ArrayOutput);
		\draw [thick,->] (ArrayOutput) -- (PatternOut);
		%%%%%%%%%%%%%%%%%%%%%% Linse %%%%%%%%%%%%%%%%%%%%%%%%%%%%%%%%%%%%
		%	\draw[thick,blue]   (5.25,0.75) rectangle ++(0.5,0.5);
		%	\draw[thick,blue]   (8.76,1.7) rectangle ++(3.5,0.1);
		%	\draw[thick,green]   (5.5,0.5) rectangle ++(0.5,0.5);
		%	\draw[thick,green]   (8.76,1.5) rectangle ++(3.5,0.1);
		%%%%%%%%%%%%%%%%%%%%%%%% text 
		%%%%%%%%%%%%%%%%%%%%%%%%%%%%%%%%%%%%%
		%\node[draw] at (0,1) {$\bm{I}$};
		\node[text width=1cm] at (3.60,-1.75) {$n\times 1$};
		\node[text width=2cm] at (6,-1.75) {$d\times(n-d+1) $};
		\node[text width=1cm] at (16.60,-1.75) {$n\times 1$};
		\node[text width=2cm] at (14,-1.75) {$d\times(n-d+1) $};
		%		\node[text width=2cm] at (8.75,-2) {$ \substack{\mathbf{Row~and~Column}\\~\mathbf{Weights}}$};
		\node[text width=1cm] at (-1,2.25) {$\mathbf{Original~Pattern} $};
		\node[text width=1cm] at (19,2.25) {$\mathbf{Estimated~Pattern} $};
		%	\node[text width=0.5cm] at (2,1.55) {$\overbrace{~~~~~~~~~~}^{\bm{D}_{n_1\times n_1}\bm{I}}$};
		%	\node[text width=0.5cm] at (3.3,-1.6) {$\underbrace{~~~~~~~~~~}_{\bm{I}\bm{D}^{\rm T}_{n_2\times n_2}}$};
		\end{tikzpicture}}
	\caption{This figure shows an overview of the proposed DOA estimation method. 
	%The input data is $\mathcal{P}_{\widetilde{\Omega}}(\mathbf{y}) \in \mathbb{C}^n$ includes $m$-element SLA data and zeros at the missing locations. It is turned into a Hankel structure $\mathscr{H}(\mathcal{P}_{\Omega}(\mathbf{y}))$. Next, the Hankel structure is completed by minimizing its nuclear norm ($\|\mathscr{H}\|_{*}$). Hankel adjoint operator is used to turn back data into normal presentation $\widehat{\mathbf{y}}$. Finally, the DOAs could be estimated from the recovered data $\widehat{\mathbf{y}}$.
	}
	\label{fig:Method}
\end{figure*}
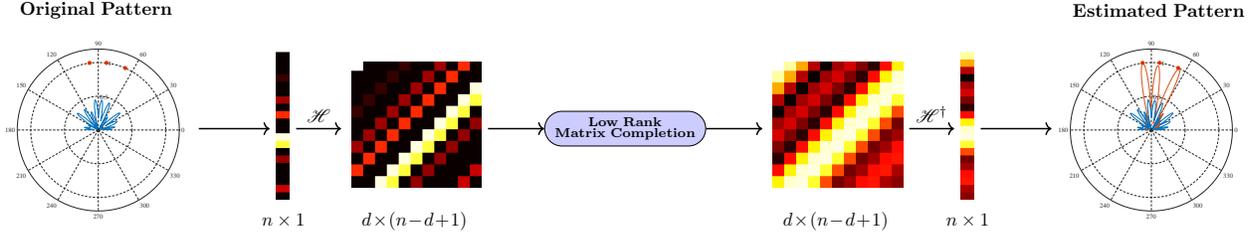

%In this paper, we proposed a novel method inspired by the EMaC method in method  \cite{chen2014robust} to interpolating the SLA array. 
% Indeed, from matrix completion prospective, to completing a matrix we need the matrix to be both low-rank and it's data not dense which it means the matrix has low leverage scores. 

%In the standard matrix completion problem\cite{candes2009exact}, we need the matrix to be both low-rank and also not dense, which leads to low leverage scores.The EMAC method amid to remove the low-rankness property, and  by lifting the input data to the higher dimension, the resulting matrix would automatically be low rank (the number of sources is usually quite smaller than the size of the array)

%{\color{edit}
%The EMaC method requires low-rank matrices, this usually achieved by increasing the dimension of input data (the number of sources is usually very smaller than the size of array),}

%
%Nevertheless, leverage scores can still be high because of uniform sampling, which can lead to more samples for perfect recovery. One possible way of mitigating this issue is to sampling non-uniformly each input element proportional to the data distribution. Accordingly, the distribution of data needs either to be known as prior knowledge or estimated from partial data. 

Besides the low-rank property of the matrix, most matrix completion techniques require the availability of uniform samples from the matrix. 
For non-uniformly sampled low-rank matrices, a two-phase 
sampling-recovery strategy is proposed in  \cite{chen2015completing} for matrix-completion. However, the method does not work for matrices with structure such as Hankel matrices.

%Besides the low-rank property of the matrix, most matrix completion techniques require the elements of the matrix not to be dense, while the Hankel structure matrix would be dense while the targets get close to each other \cite[Equation 23]{chen2014robust}. For solving the issue, one possible way is to use an adaptive sampling strategy to select elements from the dense part of the matrix like the method in \cite{chen2015completing}. However, the method does not work for matrices with a structure such as Hankel matrices.

%Besides the low-rank property of the matrix, most matrix completion techniques require the elements of the matrix not be dense while, the Hankel structure matrix would be dense while the targets get close to each other \cite[Equation 23]{chen2014robust}. To solve this issue, one possible way is to use adaptive sampling strategy to pick elements from the dense part of the matrix like the method in \cite{chen2015completing}. However, the method does not work for matrices with structure such as Hankel matrices.

In this paper, we propose a new ULA sampling scheme tailored for the EMaC method; based on the received data of the first snapshot, we estimate the amount of possible information provided by each array element, and  choose a small subset of most informative elements as the active ones for the next snapshot. Next, we employ the EMaC method to interpolate for the whole ULA and apply a standard DOA estimation technique. From simulation results, we observe that the proposed method could outperform the existing approaches.

%that choose array elements with probability which is proportional to scores that defines the importance of matrix elements in structured matrix completion. On one hand, this array selection technique improves significantly  EMaC's interpolation quality such that the estimation outperform the other methods using unifrom sampling strategy (see Sec \ref{Sec:Simulation}), and on the other hand, it aids us to use few antenna in practical scenarios and switching time(reducing the capturing time \cite{Per2005DOA,jenkins1991small}.
 
For the proposed technique, we provide a theoretical sufficient bound for the number of required samples that lead to prefect recovery; the number is proportional to the average (and not the previously shown maximum in EMaC) of the leverage scores. % and not the maximum one as in \cite{chen2014robust}. 
%In fact, t
These scores intuitively describe the amount information that each sample (array element) carry about the desired DOAs. 
%of different matrix elements (equivalently various antenna locations in ULA). 
Based on leverage scores of one snapshot, we decide how to choose array elements in the next snapshot. 
The samples obtained at array elements of the second snapshot are then interpolated 
to form a ULA. 
%Then, received values at non-sampled array elements will be interpolated (by most valuable elements) to improve the DOA estimates' accuracy. 
For the interpolation, we minimize the cost function introduced in \cite{ye2016compressive} using an alternating direction method of multipliers (ADMM) approach. 
% We use an alternating direction method of multipliers (ADMM) approach to solve the optimization proposed in \cite{ye2016compressive}. The two-snapshot version of this approach  is discussed in this paper; however, it is applicable to multi-snapshot scenarios too. 
Simulation results show that the proposed method outperforms the existing methods such as 
%Our method surpasses previous techniques such as 
ANM, EMaC, and a grid-based CS method.% based on simulation results.

%% We would need to edit it  at the end if the configuration  is finalized 
%{ \color{red} We organize the paper as follows: the system model is explained in Sec. \ref{Sec:Model}. Then, the DOA estimation approach for the two-snapshot scenario is defined in Sec. \ref{Sec:DOAestimation}. the numerical simulations are presented in \ref{Sec:Simulation}, finally the paper is concluded in Sec. \ref{Sec:Conclude}.}

\vspace{0.25cm}

\noindent
{\bf Notations:}
Lowercase boldface letters (eg. $\xv$) are used for column vectors and uppercase boldface letters (eg. $\Mv$) represent matrices. 
The set of all Hankel matrices of size $n \times d$ is shown by $\Hcal(n,d)$.
%
%Is the following, we shall often use 
The $(n,d)$-\emph{Hankelization} of a vector $\xv$ of dimension $m=d+n-1 \in \Nbb$ is defined as $\Mv=\Hscr (\xv) \in \Hcal(n,d)$ where
\ea{
\Mv_{ij} = \xv_{|i+j-1|},
}
that is, $\Hscr (\xv)$ is the Hankel matrix appropriately constructed from the elements of the vector $\xv$ by placing the $i^{\rm th}$ element in $\xv$ identically on the $i^{\rm th}$ anti-diagonal of $\Mv$.
Note the Hankelization of a vector is sometimes referred to as \emph{Hankel structure matrix}, see \cite{ye2016compressive}.
%

%\vspace{0.5cm}
$\rank(\cdot)$, , $\cdot^{\rm H}$,  $\|.\|_{\rm F}$ and $\|.\|_{*}$ stand for the rank operator, Hermitian operator (conjugate-transpose), the Frobenius norm and the nuclear-norm of a matrix, respectively. %$\mathbb{C}^{n}$.
%where $\|.\|_{*}$ stands for nuclear norm. 
Also, we use $\mathds{1}_{\mathbf{x},\Omega}$ to represent the indicator function where it equals to $\mathbf{x}_i$ for $ i  \in \Omega$ and infinity otherwise.

\section{System Model}\label{Sec:Model}
%\vspace{0.25cm}
%{\color{blue} SR2AA: thaht's not really a system model ATM}
%\vspace{0.25cm}

Suppose an array with $m$ elements which are chosen from $n$ possible uniformly spacing locations along a line ($m \leq n$). In the uniform setting, each element has  distance $d$ from  its neighbors. The phase shift of the received signal (with wavelength $\lambda$) from a source at angle $\phi$ at $k$-th antenna element  relative to the first element (the reference) is found by $a_k(\phi) = \exp{\big(-{\rm j}\tfrac{2\pi}{\lambda}kd\sin (\phi)\big)}$. Similarly, the overall received signal from sources located at angles $\phi_1,\dots,\phi_r$ is
\begin{align}
\label{eq:measurement}
y_k = \sum_{\ell=1}^{r} b_{\ell}a_{k}(\phi_{\ell}).
\end{align}
where $b_{\ell}$s are complex numbers that encode both the received power from the sources and their phase values with respect to a reference. To summarize, we define $\tau_{\ell} = \frac{d}{\lambda}\sin (\phi_{\ell})$ which converts the measurements model  \eqref{eq:measurement} into $y_k = \sum_{\ell=1}^{r} b_{\ell}{\rm e}^{-{\rm j}2\pi\tau_{\ell}k }$. Now, the DOA problem is to estimate $\tau_1,\dots,\tau_{r}$ by observing $y_{\Omega_1},\dots,y_{\Omega_m}$, where $\Omega\subset\{1,\dots, n\}$.
%$m$-selected antennas elected in $\Omega\subset[n]$.
% $y_k,~k\in\Omega$ in which $\Omega\subset[n]$ includes the location of the antennas ($[n]$ is defined as $\{1,\dots, n\}$). 
In vectorial form, we represent the full ULA samples as $\mathbf{y} = [y_1,\dots , y_n]^T$ and denote the observed (available) samples by 
%By using vector representation of measurements $\mathbf{y} = [y_1,\dots , y_n]^T$, the observation vector is 
$\mathbf{y}_{\rm o} = \mathcal{P}_{\Omega}(\mathbf{y})$ where the projection operator  $\mathcal{P}_{\Omega}$ equals to $\mathbf{y}[{k}]$ for $k \in\Omega$ and zero for the $ k \not\in \Omega$.

%\begin{align}
%\nonumber
%&\mathbf{y}_{\rm o} = \mathcal{P}_{\Omega}(\mathbf{y}) :=\left\{\begin{array}{ll}
%\mathbf{y}[{k}], & k \in\Omega,\\
%0,& k \not\in \Omega.
%\end{array}
%\right. .
%\end{align}
% 

%\section{DOA estimation}\label{Sec:DOAestimation}

\section{Proposed method}\label{Sec:DOAestimation}

%As explained earlier, we first convert the SLA into a ULA. For this purpose, we consider the Hankelization matrix of the received samples on the ULA from which we have observed only those samples that belong to the SLA; the latter matrix is low-rank and is found to be 
%particularly effective in the line spectral estimation problem \cite{razavikia2019reconstruction,razavikia2019sampling,ye2016compressive}. 
%We use the data of the first snapshot to determine the most informative array elements for matrix completion. 

%In our approach, the first step consists in converting the SLA measurement vector $\mathbf{y}_{\rm o}$ into the ULA vector $\mathbf{y}$ through a two-phase structured matrix completion algorithm. 
%
%
%In fact, we seek $m$ elements that are the most informative one in terms of the structured matrix completion.
%
%The aim of this  algorithm is to identify the $m$ matrix elements that are the most informative in terms of matrix completion.
%
%{\change 
As explained earlier, we assume that a full ULA is available, however, due to implementation and timing budgets, the data of only a few ($m$) array elements  are measured and processed. At the time of receiving the first snapshot, a predefined subset  $\Omega\subset \{1,\dots,n\}$ of   array elements are active. Based on the measured data $\mathcal{P}_{\Omega}(\mathbf{y}_1)$, we assign a score (leverage scores) to each array element in the ULA which roughly shows the amount of information that the element conveys about the targets. For the second snapshot, we activate the $m$ most significant array elements $\widetilde{\Omega}\subset \{1,\dots,n\}$ according to the scores. Next, we use EMaC to predict the data of the $n-m$ unmeasured array elements. Finally, we estimate the DOAs based on the full ULA data ($\widehat{\mathbf{y}_2}$).

The concept of leverage scores was  introduced in \cite{chen2015completing} for enhancing the performance of matrix completion in an adaptive sampling scenario. It is known that sampling the matrix elements with probabilities proportional to the leverage scores reduces the number of required samples for perfect recovery. Unfortunately, the leverage scores defined in \cite{chen2015completing} are not applicable to Hankel-structured matrices, as the elements in a Hankel matrix are repeated multiple times and cannot have independent scores. Here, we adapt the scoring concept to Hankel matrices.

%The leverage scores concept first was introduced to enhance matrix completion performance when utilizing adaptive sampling technique \cite{chen2015completing}. 
%
%
% 
%
%In Sec. \ref{sec:Theoritical}, we  show that
%the number of samples sufficient for exact matrix recovery is  decreased if each element is sampled with probability based on the leverage scores. 
%
%To utilize these scores, we need to tailor the definition  of the scores in \cite{chen2015completing} for the Hankel structure in the following definition.

\begin{definition}
	\label{LEV_SCORS}
	For an odd integer $n$ and ~$\mathbf{x} \in \mathbb{C}^{n}$, let  $\mathbf{U}_{N\times r}\boldsymbol{\Sigma}_{r\times r} (\mathbf{V}_{N\times r})^{\rm H}$ be the singular value decomposition (SVD) of the Hankelization of $\mathbf{x}$, $\mathscr{H}(\mathbf{x})$, where $N = (n+1)/{2}$ and $r = \rank(\Hscr(\xv))$.
	%Consider  $\mathscr{H}(\mathbf{x})$ for $\mathbf{x} \in \mathbb{C}^{n}$ ($n$ is an  odd integer and $N = (n+1)/{2}$) is  $r$ and let $\rank(\Hscr(\xv))=r$ and $\mathbf{U}_{N\times r}\boldsymbol{\Sigma}_{r\times r} (\mathbf{V}_{N\times r})^{\rm H}$ be the singular value decomposition (SVD) of $\mathscr{H}(\mathbf{x})$. 
	We define the leverage scores $\{\mu_{k}\}_{k=1}^{n}$ as
	\begin{align}
	\label{eq:muDefinition}
	\mu_{k} :=\frac{n}{r} \max\lcb \|\mathbf{U}^{\rm H}\mathbf{A}_{k}\|_{\rm F}^2,
	\| \mathbf{A}_{k}\mathbf{V}^{\rm H} \|_{\rm F}^2   \rcb,
	\end{align}
	for $ k \in\{1,\dots,n\}$ by setting
%	\ea{
%	\mathbf{A}_{k} := \frac{\mathscr{H}(\mathbf{e}_i^n)}{\| \mathscr{H}(\mathbf{e}_i^n) \|_{\rm F}},
%	\label{eq:A_k}
%	}
    $\mathbf{A}_{k} := \frac{\mathscr{H}(\mathbf{e}_i^n)}{\| \mathscr{H}(\mathbf{e}_i^n) \|_{\rm F}}$,
	where $\mathbf{e}_i^n $ denotes the $i$th canonical basis of $\mathbb{R}^n$.
%	$n$-dimensional basis vector. %
We should note that the set $\{\mathbf{A}_{k}\}_{k =1}^{n}$ spans the space of $N\times N$ Hankel matrices.
	
% 	\textit{which $i$-th element is one ($\mathbf{A}_{k}$s span the Hankel matrices of size $N\times N$).}
\end{definition}
%
%{ \change
%Note that the set $\{\mathbf{A}_{k}\}_{k \in [n]}$ in \eqref{eq:A_k} spans the Hankel matrices of size $N\times N$.
%}
%
%Therefore, based on this definition, how much each vector's elements are important to be sampled.
%

Since the data of the full ULA is not available, we need to approximate  the leverage scores in Def. \ref{LEV_SCORS} from the SLA associated with $\Omega$. 
%Due to the inaccessibility of full ULA, we need to approximate  the leverage scores in Def. \ref{LEV_SCORS} from SLA data in the first snapshot $\mathbf{y}_{1}$.
%
%To to this, it only requires to obtain the SVD of the
This approximation can be obtained from the SVD of 
$\mathscr{H}\big(\mathcal{P}_{\Omega}(\mathbf{y}_{1})\big)$ i.e. $\widetilde{\mathbf{U}}\widetilde{\boldsymbol{\Sigma}} (\widetilde{\mathbf{V}})^{\rm H}$ for computing the approximate leverage scores $\widetilde{\mu}_k$, which determine the active array elements  
%to choose new locations set 
$\widetilde{\Omega}$ for the next snapshot.

Once the set $\widetilde{\Omega}$ has been identified, we consider the following matrix completion problem to interpolate values at missing locations:
%=======================================
\begin{equation}
\label{convex-opt}
\widehat{\mathbf{y}}_2 = \underset{ \mathbf{g} \in \mathbb{C}^{n}}{\rm argmin}~~~
 \| \mathscr{H}({\mathbf{g}}) \|_{*}, \ \  {\rm s.t.}~~~
  \mathcal{P}_{\widetilde{\Omega}}( \mathbf{g})  = \mathcal{P}_{\widetilde{\Omega}}( \mathbf{y}_2).
\end{equation}
%=======================================
The optimization problem in \eqref{convex-opt} is a convex relaxation of matrix rank minimization. 
In order to reduce the computational complexity of the minimization, we solve the nuclear norm minimization \eqref{convex-opt} using a SVD-free method. 
As shown in  \cite{srebro2004learning}, we can replace the nuclear norm of a matrix $\mathbf{A}$, i.e. $\|\mathbf{A}\|_{*}$, with the $ \underset{\mathbf{U},\mathbf{V}}{\min}\| \mathbf{U} \|_{\rm F}^2+\| \mathbf{V} \|_{\rm F}^2$ subject to $\mathbf{A}=\mathbf{UV}^{\rm H}$. 
%=======================================
%\begin{align}
%\| \mathbf{A} \|_{\rm *} = \underset{\substack{\mathbf{U},\mathbf{V}\\ \mathbf{A}=\mathbf{UV}^{\rm H}}}{\min}\| \mathbf{U} \|_{\rm F}^2+\| \mathbf{V} \|_{\rm F}^2.
%\end{align} 
%=======================================
As a result, \eqref{convex-opt} can be represented as
%=======================================
\begin{equation*}
\label{eq:minUV}
\underset{ \mathbf{U},\mathbf{V} ,\mathbf{g} \in \mathbb{C}^n}{\rm argmin}~ \| \mathbf{U} \|_{\rm F}^2+\| \mathbf{V} \|_{\rm F}^2 
~{\rm s.t.}~\mathcal{P}_{\widetilde{\Omega}}(\mathbf{g})  =\mathcal{P}_{\widetilde{\Omega}}( {\mathbf{y}}),{\rm and}~ \mathbf{g} =\mathbf{UV}^{\rm H}.
\end{equation*}
%=======================================
Next, we combine the two constraints to apply the ADMM method in the following cost function:
\begin{align}
\label{eq:ADMM_equation}
L_{\rho}(\mathbf{U},\mathbf{V}, &\mathbf{g},\mlambda) = \| \mathbf{U} \|_{\rm F}^2+\| \mathbf{V} \|_{\rm F}^2 + \mathds{1}_{\mathbf{g},\Omega}\nonumber\\ 
& + \rho \|\mathscr{H}(\mathbf{g}) -\mathbf{UV}^{\rm H}+ \mlambda \|_{\rm F}^2, 
\end{align}
where $\mathds{1}_{\mathbf{g}, \Omega}$ denotes the indicator function.
%where $i(\mathbf{g})$ denotes the indicator function which is $0$ for $ \mathcal{P}_{\widetilde{\Omega}}(\mathbf{g}) = \mathcal{P}_{\widetilde{\Omega}}( {\mathbf{y}})$ and infinity otherwise.
%{\color{blue} SR2all:use $\ones$ for the indicator function,put in notation section.}
%\begin{align*}
%i(\mathbf{g}) =  \left\{\begin{array}{lc}
%0, & \mathcal{P}_{\widetilde{\Omega}}(\mathbf{g}) = \mathcal{P}_{\widetilde{\Omega}}( {\mathbf{y}}),\\
%\infty, & \textrm{otherwise}.
%\end{array}
%\right.
%\end{align*}
Also, $\rho$ is a positive scalar and the Lagrange multiplier $\mlambda$ has the same size as $\mathscr{H}(\mathbf{g})$. 
To implement the ADMM technique, we shall minimize $L_{\rho}(\mathbf{U},\mathbf{V}, \mathbf{g},\mlambda)$ sequentially in terms of its argument in turn. We skip the details here. 
%it only needs to take the derivation of each variable in Lagrange cost function separately. 
Although minimizing $L_{\rho}$ is not a convex program, the  analysis in \cite{hong2016convergence} guarantees the convergence  when the penalty parameter $\rho$ is sufficiently large. Finally, the DOAs can be estimated using standard super-resolution methods, see \cite{prony1795essai}.

% =======================================================
%########################################################
% =======================================================

%\subsection{DOA estimation for  Uniform Linear Array}

%Until now,we can reach a fine set of array elements in a few snapshots and obtain a full uniform array $\widehat{\mathbf{y}}_2$ by interpolating a non-uniform array $\mathcal{P}_{\widetilde{\Omega}}(\mathbf{y}_2)$. In the next step, to estimate the DOAs, we apply the super-resolution technique. To solve this problem, we used the proposed approach in \cite{candes2014towards} and summarized it.

% ========================================
\begin{algorithm}[tb]
\caption{Hankel interpolation using ADMM}\label{alg:Low-rank-recovery}
\begin{algorithmic}[1]
	\State \textbf{Input:}
	\State {\quad Observed samples $\mathcal{P}_{\Omega}(\mathbf{y}_1)\in \mathbb{C}^{n}$ and corresponding coordinates $ \Omega \subset [n] $},
	\State {\quad Parameters ${\rho}$ for the augmented Lagrangian form }
%		\State \quad The number of iterations $K$,
	\State \textbf{Output:}
	\State {\quad Completed Vector }$\widehat{{\mathbf y}}_{2} \in \mathbb{C}^{n}$.
	\Procedure{Hankel\_ Interpolation}{${\mathcal{P}_{\Omega}(\mathbf{y}_1)},{\rho}$}
	%\State $ i = 1,  $
	\State Estimate leverage scores $\mu_{k}$ by \eqref{eq:muDefinition} using $\mathcal{P}_{\Omega}(\mathbf{y}_1)$ 
	\State Construct $\widetilde{\Omega}$ using $\mu_{k}$ and change sampled locations
	\State Get new samples $\mathcal{P}_{\widetilde{\Omega}}(\mathbf{y}_2)$
	\State Solve ADMM problem in \eqref{eq:ADMM_equation} (consider $\rho$ as its input) by proposed method in \cite{ye2016compressive} to find full array data $\widehat{\mathbf{y}}_2$
	\State \textbf{return} $\quad\widehat{\mathbf{y}}_2$
	\EndProcedure
\end{algorithmic}
\end{algorithm}
\begin{figure*}[!t]
	\centering
		\begin{minipage}{.23\linewidth}
			\centering
			\includegraphics[width=1\linewidth]{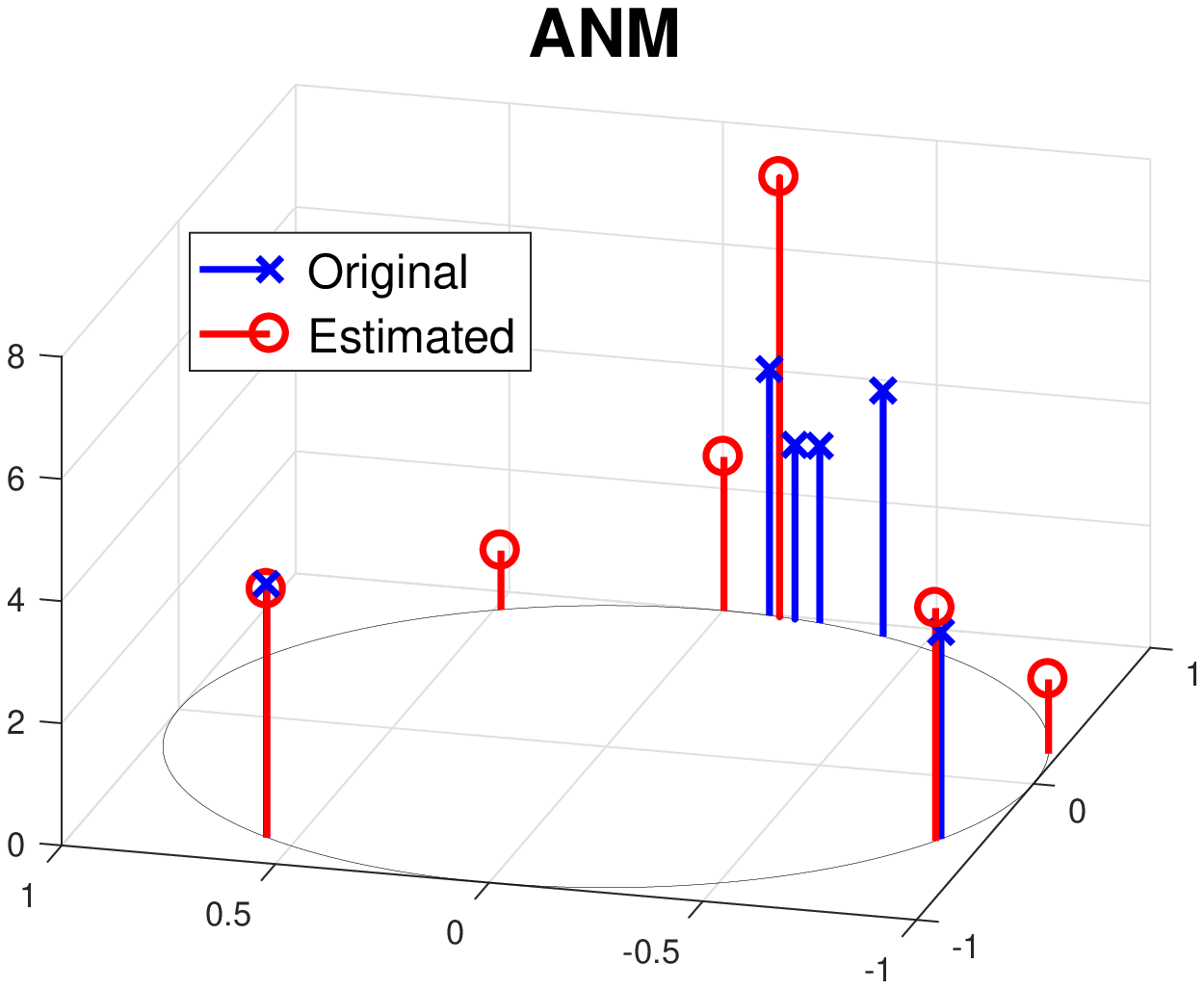}
			(a)
		\end{minipage}
		\begin{minipage}{.23\linewidth}
			\centering
			\includegraphics[width=1\linewidth]{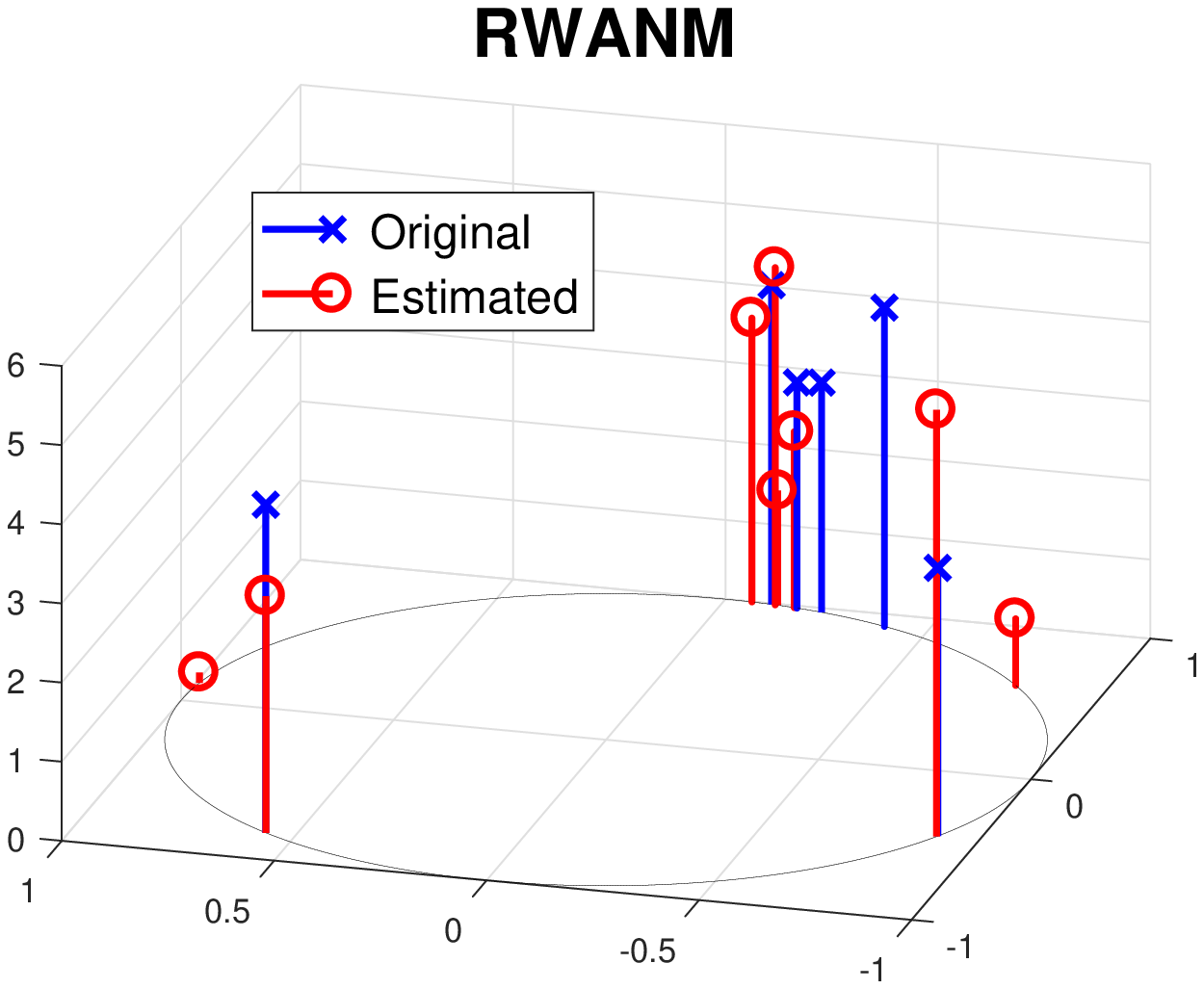}
		    (b)
		\end{minipage} 
		\begin{minipage}{.23\linewidth}
			\centering
			\includegraphics[width=1\linewidth]{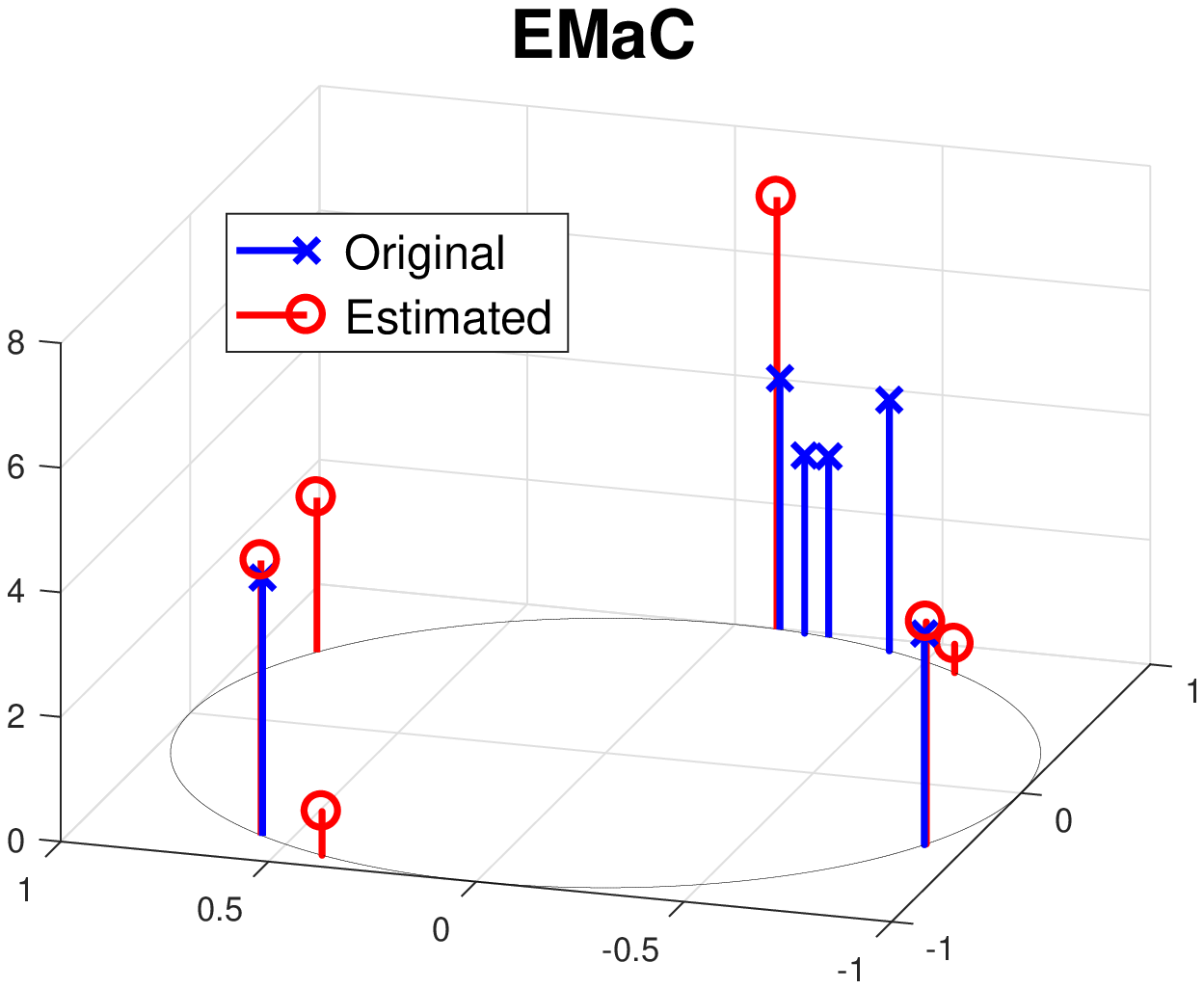}
	        (c)
		\end{minipage} 
%end{figure*}
%begin{figure*}[!t]
        %\begin{minipage}{.24\linewidth}
	%		\centering
	%		\includegraphics[width=1\linewidth]{figs/snapshot1/cmra.eps}
	%		(d)
	%	\end{minipage}
        %\begin{minipage}{.24\linewidth}
		%	\centering
		%	\includegraphics[width=1\linewidth]{figs/snapshot1/rwAnm.eps}
		%	(e)
		%\end{minipage} 
		\begin{minipage}{.23\linewidth}
			\centering
			\includegraphics[width=1\linewidth]{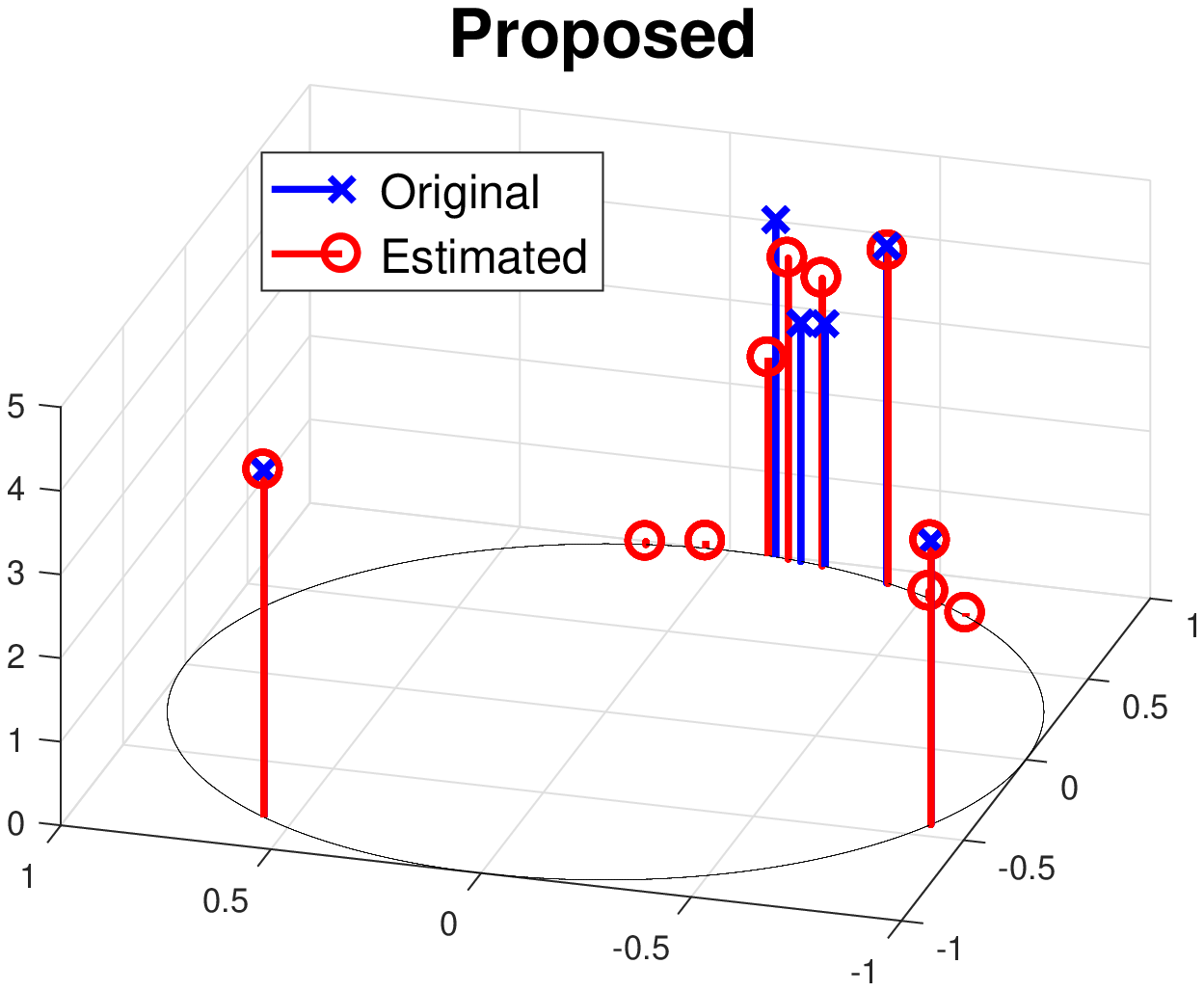}
			(f)
		\end{minipage} 
	% ---------------------------------------
	\caption{Comparison of the performance in DOA estimation; the height of the bars shows the received signal power, and base circle in each figure represents the $360^{\circ}$ angular space. Here, we have $6$ sources with one almost collocated triplets.}
	\label{fig:locations}
	
\end{figure*}
% ========================================

\section{Guarantee for prefect recovery}\label{sec:Theoritical}

Below, we provide a theoretical result for the number of required samples. For this purpose, we use the simplifying assumption that the sampled set $\Omega$ is formed by selecting $i\in\{1,\dots,n\}$ with probability $p_i$ independently of $j\neq i$. We now show that if $p_i$s are large enough, we can correctly recover the DOAs.

%Our array interpolation method depends on structured nuclear norm minimization. Here, we guarantee the conditions under which the optimization can uniquely find the DOAs. In random sampling scenario, sampled set $\Omega$ is formed by selecting $i\in[n]$ with probability $p_i$ independently of other elements $j\neq i$. In what follows, we provide a set of lower-bounds on $p_i$s to guarantee perfect recovery with high probability.

%\begin{theorem}\label{th:recovery}
%	Let $\mathbf{y}\in\mathbb{C}^{n}$ represent the hypothetical samples of the ULA for an $r$-target radar scene. 
 %   If the index of the existing antenna elements in the SLA are randomly chosen by uniformly drawing the indices from $\{1,\dots, n\}$, we can recover $\mathbf{y}$ from the measurements on the SLA $ \mathcal{P}_{\widetilde{\Omega}}(\mathbf{y})$ using \eqref{convex-opt} with  probability no less than $1-n^{3-b_1}$ if
%\begin{align}\label{eq:sample_com_exact}
%p_k \geq \min\Big\{ 1 ~,~ \frac{\max\big( c{\mu}_{k}r^2\log^3{(n)} \,\,,\,\, 1\big)}{n} \Big\}, 
%\end{align}
%and $\frac{1}{8\log(n)}\leq\min\{\|\mathbf{U}\mathbf{U}^{\mathsf{H}}\mathbf{e}_1^d\|_{\rm F}^2,\|\mathbf{e}_n^{n-d+1}\mathbf{V}\mathbf{V}^{\mathsf{H}}\|_{\rm F}^2\},$ 
%where $d$ is the pencil parameter used in the Hankel operator and $\mathbf{U}, \mathbf{V}$ are the unitary matrices in the SVD of $\mathscr{H}(\mathbf{y})$. Also $\mu_{k}$ is the leverage score in Definition \ref{LEV_SCORS} and $c =  192^2(b_1+1)$ for $b_1 > 3$. 
%\end{theorem}

\begin{theorem}\label{th:recovery} 
	Let $\mathbf{y}\in\mathbb{C}^{n}$ be the vector of true samples of the ULA for an $r$-sources. 
	%If the index of the existing antenna elements in the SLA are randomly chosen by uniformly drawing the indices from $[n]$, we can recover $\mathbf{y}$ from the measurements on the SLA $\mathcal{P}_{\widetilde{\Omega}}(\mathbf{y})$
	$\mathbf{y}$ can be recovered with  probability no less than $1-n^{-10}$
	from the measurements on the SLA $ \mathcal{P}_{\widetilde{\Omega}}(\mathbf{y})$
	where $\Omega$ i.e. index set of the location of the antenna elements in the SLA are randomly chosen by uniformly drawing the indices from $[n]$ by  
	solving the optimization in  \eqref{convex-opt} if
\begin{align}\label{eq:sample_com_exact}
p_k \geq \min\Big\{ 1 ~,~ \frac{\max\big( c{\mu}_{k}r^2\log^3{(n)} \,\,,\,\, 1\big)}{n} \Big\}, 
\end{align}
and $\frac{1}{8\log(n)}\leq\min\{\|\mathbf{U}\mathbf{U}^{\mathsf{H}}\mathbf{e}_1^d\|_{\rm F}^2,\|\mathbf{e}_n^{n-d+1}\mathbf{V}\mathbf{V}^{\mathsf{H}}\|_{\rm F}^2\},$ 
where $d$ is the pencil parameter used in the Hankel operator and $\mathbf{U}, \mathbf{V}$ are the unitary matrices of SVD of $\mathscr{H}(\mathbf{y})$. Also $\mu_{k}$ is the leverage score of Definition \ref{LEV_SCORS} of the paper and $c>0$ is a universal constant.\footnote{Due to the page limit, the proof of this result is provided at \href{http://sharif.ir/~aamini/Papers/Hankel_Proof.pdf}{http://sharif.ir/$\sim$aamini/Papers/Hankel\_Proof.pdf}}
\end{theorem} 
%The proof can be found \href{http://sharif.ir/~aamini/Papers/Hankel_Proof.pdf}{here}. 
As the expected number of total samples is given by $\sum_{k}p_k$, 
Theorem \ref{th:recovery} shows that the number of required samples for prefect recovery is proportional to the $\frac{\sum_{k=1}^n{\mu_k}}{n}$ instead of $\max_k \mu_k$ in \cite{chen2014robust}.

\section{Simulations}\label{Sec:Simulation}
In this section, we compare the proposed sampling algorithm with some off-the-shelf single-snapshot DOA estimation methods like  ANM \cite{bhaskar2013atomic}, RWANM \cite{yang2015enhancing} and  EMaC  \cite{chen2014robust}.
% ========================================
\begin{figure*}[ht!]
	\centering
	\scalebox{0.9}{
	\begin{minipage}{.49\linewidth}
		\centering
		\includegraphics[clip, width=1\linewidth]{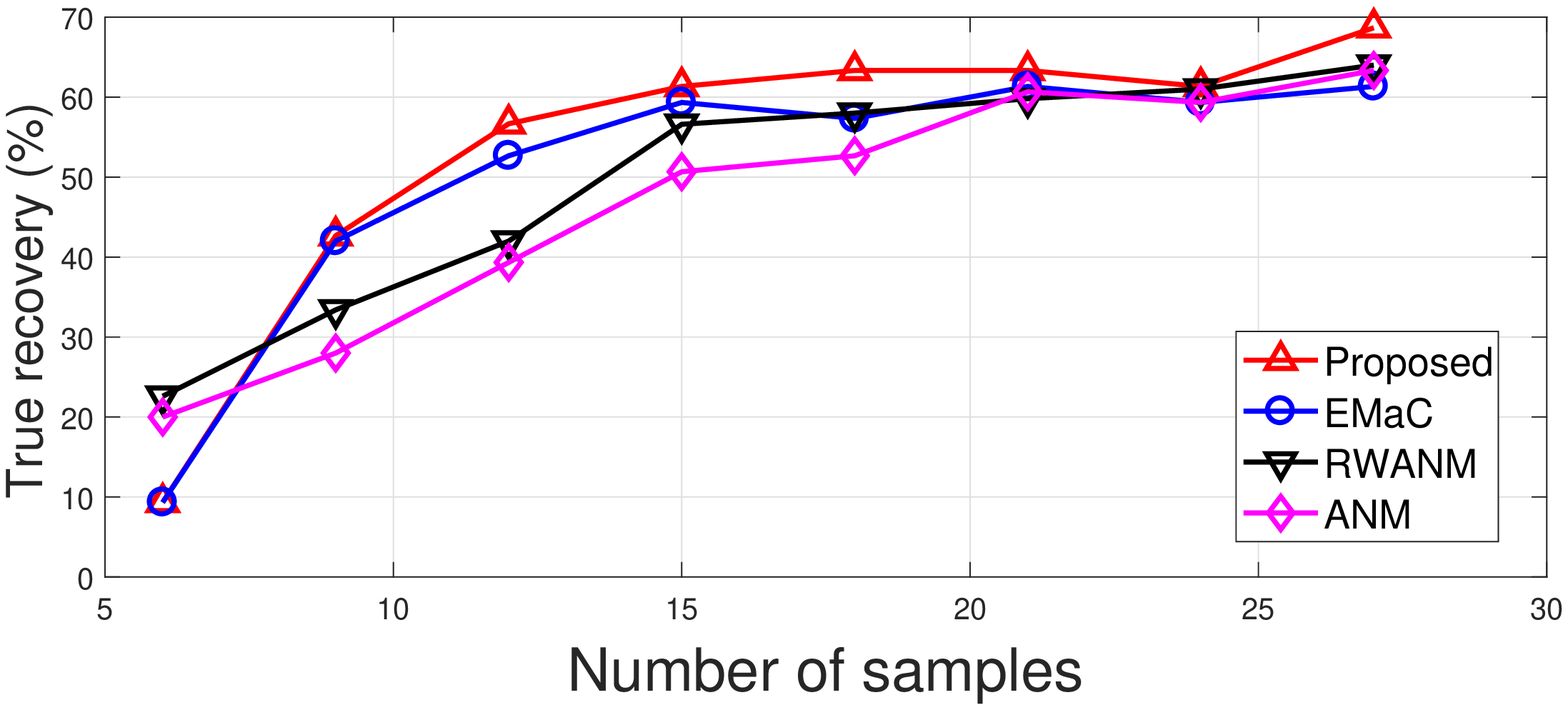}
		(a)
	\end{minipage}
	\begin{minipage}{.49\linewidth}
		\centering
		\includegraphics[width=1\linewidth]{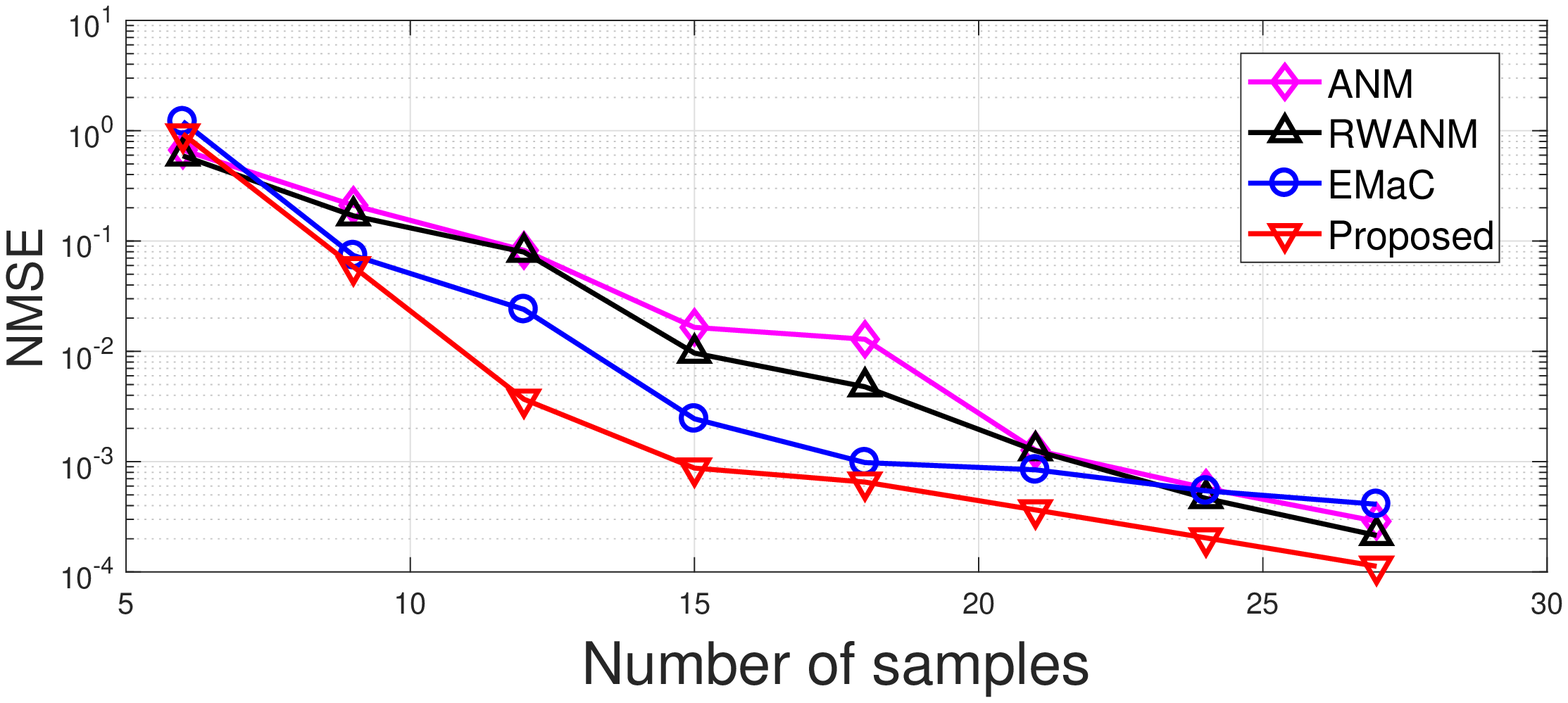}
		(b)
	\end{minipage}
	}
    \caption{Effect of the array size. The SLA interpolation normalized mean square errors are depicted for different array sizes. The source setup includes $5$ sources with two co-angled pairs.}
	\label{fig:accuracy}
\end{figure*}
% ========================================
Note that all methods, like the proposed one, inherently include an array completion procedure.
After this first phase,  subspace methods like Prony \cite{prony1795essai} and super-resolution \cite{candes2014towards} are employed to estimate the DOAs.
%
%on the completed array DOAs are recovered. 
%

%{\change  
We assess the effect of the proposed non-uniform array selection for EMaC with standard uniform sampling in terms of array interpolation and DOAs estimation. To show the improvement, we also compare the modified EMaC with the other grid-less algorithms like ANM and RWANM that outperform standard EMaC. Moreover, to fairly assess array selection, the algorithms only used one snapshot of data to estimate the DOAs. In fact, we want to assess the effectiveness of the array selection in comparison with uniformly at random selection. Note that this strategy is only designed for the Hankel structure in the EMaC algorithm.%, and it is not suited to be applied for the other algorithms.
Furthermore, we use only one snapshot for estimating the leverage scores; however, it is possible to increase this number in practical scenarios to alleviate system imperfections such as noise.
%we used the only first snapshot for estimating the leverage scores; however, in real-time scenarios, the first snapshot's adverse effect is decreased by the number of snapshots increasing.}

%{\change We should note that, in the simulations for all methods, only the second snapshot data are used to recover ULA and estimate the DOAs. The first snapshot data only is used in proposed method at the array selection estimation step. At the first look, it seems unfair comparison between methods because more data is used for the proposed method. However, in this paper, we assess the effect of array selection and comparing methods on single snapshot data is better way than two snapshot case to evaluate this effect. Because in two snapshot case, the first snapshot is not chosen based on leverage scores and this undesired array selection in first snapshot diminish real potential of the proposed array selection method. Moreover, in real-time scenarios by number of snapshot increasing the adverse effect of the first snapshot is decreased. }

In the experiments, 
%Also, we suppose 
the ULA elements are placed on integer multiples of $\lambda/2$,  where $\lambda$ is the wavelength.
In the first snapshot, we randomly select $\Omega$ (active elements) %the SLA elements from ULA. 
%
%{\change 
and use the samples for estimating the leverage scores defined in \eqref{eq:muDefinition}. 
Before the second snapshot, we activate ULA elements 
%In the second snapshot, we non-uniformly select the array elements for 
with probabilities proportional to the estimated  leverage scores. %as a proposed scenario for the EMaC. 
For other methods, ULA elements are activated with uniform probability. 
%Furthermore, for the remaining algorithms, locations are selected randomly with uniform probability
%In the second snapshot, we have two different approaches: (i) for the proposed method, the SLA elements are selected randomly by leverage scores imposing probability  which are calculated from the first snapshot's data, while (ii)
%for the remaining algorithms, locations are selected randomly with uniform probability.
%} 
%
% In the proposed method for the next snapshot, SLA elements are selected based on calculated leverage scores in the previous snapshot; however, locations are selected randomly for other algorithms. 
%
%
%Moreover, for the grid-based method we uniformly divide interval $[-1,1]$ into $2^{12}$ parts.
%
%We used the BP method to solve the standard compressed sensing problem.
%
%The results for BP
%
%In the figures, its results are shown by the label BP. 
In the ADMM algorithm, the value of $\rho$ should be sufficiently large to guarantee the convergence: in our simulations we choose $\rho = 10^{3}$.
Next, we consider two simulation scenarios: 

\noindent
%\subsection{DOA estimation}
{\bf (i) Fig. \ref{fig:locations} -- DOA estimation:} 
%\vspace{1cm}
%In the first experiment, 
we consider an array with an aperture equal to $51\lambda/2$ ($52$ elements) and activate $12$ elements in each snapshot. 
%
%In Fig. \ref{fig:locations}, we evaluate 
%the performance of our proposed method with $3$ methods (EMaC, CMRA, ANM, RWANM, and BP). 
%
In Fig.  \ref{fig:locations}, both the original DOAs and the estimated ones are plotted. 
This scenario includes $6$ sources from which $3$ are almost  collocated. 
%
%Our proposed method outperformed others. 
%First of all, we should 
%
As shown in this figure, existing methods are unable to distinguish or detect seemingly collocated sources correctly.%, and all of them find some sources with considerable amplitude in wrong  directions. 
The proposed method, however,   resolves all source locations correctly, although  with some false-positive sources with very low amplitudes.

\noindent
{\bf(ii) Fig. \ref{fig:accuracy} -- effect of the array size:} In the second experiment, we aim to evaluate the overall performance of the proposed algorithm  against other methods for different number of active array elements. %,  i.e. number of available receivers. 
As before, we consider a $51\lambda/2$ aperture size for the uniform array; further, we uniformly select $6$ to $27$ elements for the SLA while we have $6$ sources at angles  $\{-23.80^{\circ}$,$15.60^{\circ}$,$16.20^{\circ}$, $-17.53^{\circ}$,$18.13^{\circ}\}$ with amplitudes $\{3.31$,$3.2$,$2.13$,$3.14$,$3.56\}$, respectively. 
We call a detection correct, if for the source angle $\theta$ and its estimated version $\hat{\theta}$ we have that $|\sin{\theta} - \sin{\hat{\theta}}|\leq 0.005$.
%{ \change We suppose a source with angle $\theta$ is recovered if only one estimated source with angle $\hat{\theta}$ is existed that satisfies $|\sin{\theta} - \sin{\hat{\theta}}|\leq 0.005$.  }
%
In Fig. \ref{fig:accuracy}-(a),  we plot the rate of correct source recovery as a function of the SLA size (active elements). 
For each curve, the average results over $100$ random realization of the SLA elements are reported. 
%
%{\change 
We can see that the performance of the proposed sampling strategy outperforms the standard EMaC method, as well as other uniform algorithms such as ANM and RWANM.   
The recovery rate also increases as the number of the elements increase.%; this suggests that if we use more snapshots  for estimating the leverage scores estimation step, then the performance of the proposed scenario would be still better than uniform sampling strategy.
%The proposed array selection has better performance in compare to EMaC. while the ANM and RWANM have slightly better performance than EMaC for some number of samples, using the proposed sampling strategy help EMaC to outperform them.}
%}
%While the proposed method and EMaC outperform others, the proposed one has slightly better performance.
%
Fig. \ref{fig:accuracy}-(b) depicts the normalized MSE of the estimated ULA's missed elements in the SLA. 
%
%The BP method absents here because these curves involve matrix completion procedure. 
%
Almost for any array size, the proposed method has the smallest NMSE value. 

\section{Conclusion}\label{Sec:Conclude}
%We studied a two-snapshot  DOA estimation problem based on SLA data. % Introduciton ??
We studied a new ULA sampling method based on leverage scores in the two-snapshot DOA estimation problem. We proposed an approach that initially determines optimal array elements for matrix completion using the received data from the first snapshot. Next, it applies the EMaC method on the second snapshot data to find the DOAs. The sampling method is numerically observed to outperform EMaC, and other state-of-the-art grid-less DOA estimation techniques which have better performance than EMaC. besides, we provided a theoretical guarantee for perfect estimation.

%We proposed an approach that initially interpolates the SLA into a ULA through matrix completion. Next, it applies the Prony method to find the DOAs. The received data from the first snapshot determine the optimal array elements for matrix completion, while the interpolation (matrix completion) is applied to the second snapshot. 

%Our matrix completion method used the first snapshot to estimate the appropriate array location for matrix completion. Then, it completed the second snapshot by using samples of the estimated array locations.besides, we provide theoretical guarantee for the unique recovery. Based on the simulation results, the proposed method outperformed other existing grid-less methods in terms of DOA estimation accuracy and array interpolation.

\bibliography{refs}
\bibliographystyle{IEEEtran}

\end{document}